\documentclass[twocolumn,showpacs,preprintnumbers,amsmath,amssymb,superscriptaddress]{revtex4}

\usepackage{graphicx}
\usepackage{bm}

\newcommand{\conepm}{X_{9,1}}
\newcommand{\ctwop}{X_{9,2}}
\newcommand{\cfourp}{X_{9,4}}

\begin{document}


\preprint{SFB/CPP-09-19, TTP09-05}
\title{Quark and gluon form factors to three loops}
\author{P.A. Baikov}
\affiliation{Skobeltsyn Institute of Nuclear Physics of Moscow State University, 119992 Moscow, Russia}
\affiliation{Institut f\"ur Theoretische Teilchenphysik, Universit\"at Karlsruhe (TH), D-76128 Karlsruhe, Germany}
\author{K.G. Chetyrkin}
\affiliation{Institut f\"ur Theoretische Teilchenphysik, Universit\"at Karlsruhe (TH), D-76128 Karlsruhe, Germany}
\affiliation{Institute for Nuclear Research, Russian Academy of Sciences, Moscow 117312, Russia}
\author{A.V. Smirnov}
\affiliation{Scientific Research Computing Center, Moscow State University, 119992 Moscow, Russia}
\affiliation{Institut f\"ur Theoretische Teilchenphysik, Universit\"at Karlsruhe (TH), D-76128 Karlsruhe, Germany}
\author{V.A. Smirnov}
\affiliation{Skobeltsyn Institute of Nuclear Physics of Moscow State University, 119992 Moscow, Russia}
\affiliation{Institut f\"ur Theoretische Teilchenphysik, Universit\"at Karlsruhe (TH), D-76128 Karlsruhe, Germany}
\author{M. Steinhauser}
\affiliation{Institut f\"ur Theoretische Teilchenphysik, Universit\"at Karlsruhe (TH), D-76128 Karlsruhe, Germany}
\date{Februar 13, 2009}
\begin{abstract}
We compute the form factors of the photon-quark-anti-quark vertex and the
effective vertex of a Higgs boson and two gluons
to three-loop order within massless perturbative Quantum
Chromodynamics. These results provide building blocks for many third-order
cross sections. Furthermore, this is the first calculation of complete
three-loop vertex corrections.
\end{abstract}


\maketitle


In the recent years various next-to-next-to-leading order (NNLO)
calculations to physical observables have been completed.
Among them are the total threshold cross section for top quark pair
production in electron positron annihilation~\cite{Hoang:2000yr},
the Higgs boson production in gluon
fusion~\cite{Harlander:2002wh,Anastasiou:2002yz,Ravindran:2003um},
the rare decay rate of the $B$ meson into a meson containing a strange quark
and a photon~\cite{Misiak:2006zs,Misiak:2006ab} and the three-jet
cross section at lepton colliders~\cite{GehrmannDeRidder:2007hr,Weinzierl:2008iv}.
There exist also a few results at next-to-next-to-next-to-leading order
(NNNLO), like the total hadronic cross section in electron positron
annihilation~\cite{Baikov:2008jh}, the hadronic $\tau$
lepton~\cite{Baikov:2008jh} and Higgs boson
decay~\cite{Baikov:2006ch}. It is common to all NNNLO results that the calculation can
be reduced to two-point functions and that only one mass scale is involved in
the computation.

\begin{figure}[tb]
  \begin{center}
    \begin{tabular}{ccc}
      \includegraphics[width=7em]{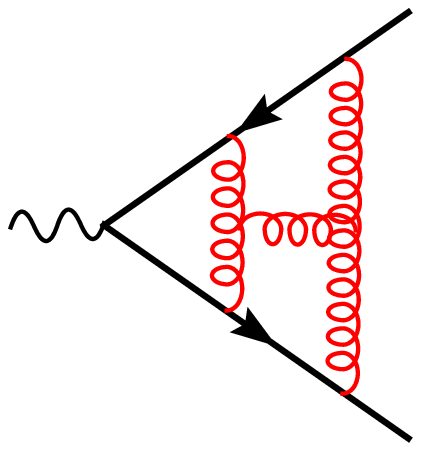} &
      \includegraphics[width=7em]{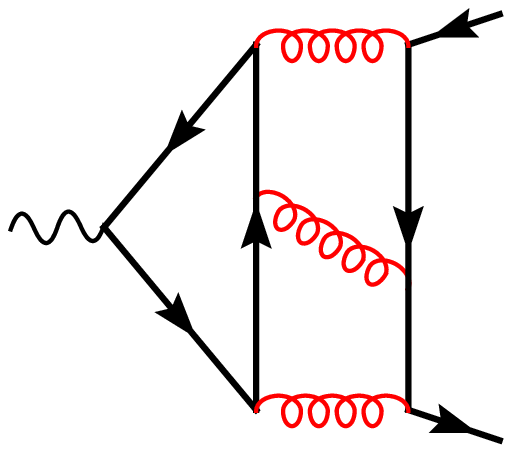} &
      \includegraphics[width=7em]{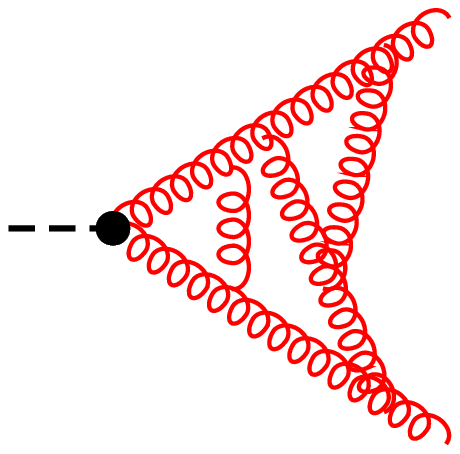}
      \\
      (a) & (b) & (c)
    \end{tabular}
    \caption{\label{fig::diags}Sample Feynman diagrams contributing to the
      $F_q$ ((a) and (b)) and $F_g$ (c)
      at three-loop order. Straight and curly lines denote
      quarks and gluons, respectively.}
  \end{center}
\end{figure}

In this Letter we provide the first NNNLO calculation of a three-point
function within Quantum Chromodynamics (QCD). To be precise, we consider
gauge invariant building blocks for NNNLO cross
sections, namely the virtual third-order corrections
for the hadronic Higgs boson production and the process
$e^+e^-\to 2$~jets. The results are conveniently expressed in terms of form
factors of the photon-quark and the effective gluon-Higgs boson vertex
originating from integrating out the heavy top-quark loops.
Denoting the corresponding vertex functions by $\Gamma^{\mu}_q$ and
$\Gamma^{\mu\nu}_g$, respectively, the scalar form factors are obtained via
\begin{eqnarray}
  F_q(q^2) &=& -\frac{1}{4(1-\epsilon)q^2}
  \mbox{Tr}\left( q_2\!\!\!\!\!/\,\,\, \Gamma^\mu q_1\!\!\!\!\!/\,\,\,
    \gamma_\mu\right)
  \,,
  \nonumber\\
  F_g(q^2) &=&
  \frac{\left(q_1\cdot q_2\,\,
      g_{\mu\nu}-q_{1,\mu}\,q_{2,\nu}-q_{1,\nu}\,q_{2,\mu}\right)}
  {2(1-\epsilon)}
  \Gamma^{\mu\nu}_g
  \,,
\end{eqnarray}
where $d=4-2\epsilon$ is the space-time dimension,
$q=q_1+q_2$ and $q_1$ ($q_2$) is the incoming (anti-)quark momentum in
the case of $F_q$, and $F_g$ depends on the gluon momenta $q_1$ and $q_2$ with
polarization vectors $\varepsilon^{\mu}(q_1)$ and $\varepsilon^{\nu}(q_2)$.
Some sample Feynman diagrams contributing to $F_q$ and $F_g$ are shown in
Fig.~\ref{fig::diags}. Starting from three-loop order a new class of
diagrams occurs, the so-called singlet diagrams, where the external photon is not
connected to the fermion line involving the final-state quarks (see
Fig.~\ref{fig::diags} (b)).
Since at three-loop order there are no counterterm contributions to the
singlet diagrams and furthermore there is no corresponding real emission contribution
the sum of all diagrams has to be finite. This
constitutes an important check on the correctness of our result.

In the recent years the evaluation of the three-loop form factor has attracted
much attention. After the pioneering work more than 20 years
ago~\cite{Kramer:1986sg,Matsuura:1987wt,Matsuura:1988sm}
where the quark form factor has been computed to two-loop order
the corresponding quantity for the Higgs-gluon coupling has been evaluated by
Harlander in Ref.~\cite{Harlander:2000mg} (see
also~\cite{Ravindran:2004mb}). The latter constitutes a building block for the
NNLO predictions of the Higgs boson production in gluon fusion at the Fermilab
Tevatron and CERN Large Hadron
Collider~\cite{Harlander:2002wh,Anastasiou:2002yz,Ravindran:2003um}.
More recently, in Ref.~\cite{Gehrmann:2005pd} the two-loop results have been reconsidered
and more terms in the $\epsilon$-expansion have been added in order to match the
three-loop accuracy. Furthermore, in
Refs.~\cite{Gehrmann:2006wg,Heinrich:2007at} almost all master
integrals necessary for the three-loop calculation have been
evaluated. However, the most complicated master integrals are still unknown.

First steps towards three-loop results for the form factors have been
undertaken in the Refs.~\cite{Moch:2005id,Moch:2005tm} where the pole parts of
$F_q$~\cite{Moch:2005id} and $F_g$~\cite{Moch:2005tm} have been
extracted from the behaviour of the three-loop
coefficient function for inclusive deep-inelastic scattering~\cite{Vermaseren:2005qc}.
Furthermore, in Ref.~\cite{Moch:2005tm} also the finite part of the
fermionic contribution to $F_q$ could be evaluated.
With our calculation we were able to confirm these results but also add the
finite contributions which are necessary for the physical observables.

For the evaluation of the Feynman integrals we developed two independent
set-ups which have in common that a reduction of all occurring integrals to
so-called master integrals is performed in $d$ space-time
dimensions. Afterwards the ($\epsilon$-expanded) master integrals are inserted.

Following Refs.~\cite{Baikov:1996rk,Baikov:2000jg,Baikov:2005nv}
one considers integral representations of the coefficient functions
of the individual master integrals in the limit of
large space-time dimension $d$, evaluates several expansion
terms and reconstructs in this way the complete rational
dependence on $d$. The most CPU-consuming step, the large $d$ expansion,
has been performed by a program written in
{\tt ParFORM}~\cite{Tentyukov:2004hz,Tentyukov:2006pr}, the parallel
version of the computer algebra program {\tt FORM}~\cite{Vermaseren:2000nd}.
For the singlet contribution, which involves
the most complicated integrals, also a second approach has been employed.
After generating the Feynman diagrams with
the help of {\tt QGRAF}~\cite{Nogueira:1991ex} they are further processed with
{\tt q2e} and {\tt exp}~\cite{Harlander:1997zb,Seidensticker:1999bb} where a
mapping to the underlying family of the diagrams
is achieved. In a next step the reduction
of the integrals is performed with the program package {\tt FIRE}~\cite{FIRE}
which implements a combination of the Laporta algorithm~\cite{Laporta:1996mq}
and a generalization~\cite{S2} of the Buchberger
algorithm (see, e.g., Ref.~\cite{Buch}) to construct Gr\"obner bases.

\begin{figure}[t]
  \begin{center}
    \begin{tabular}{ccc}
      \includegraphics[width=7em]{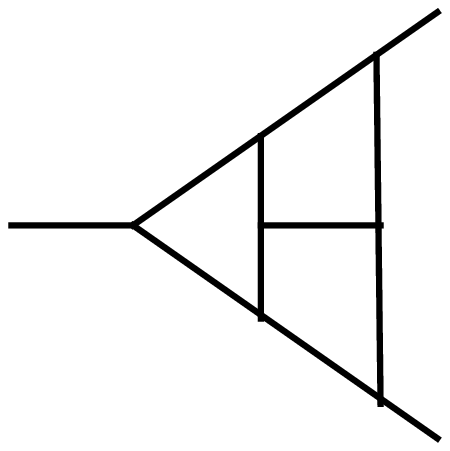} &
      \includegraphics[width=7em]{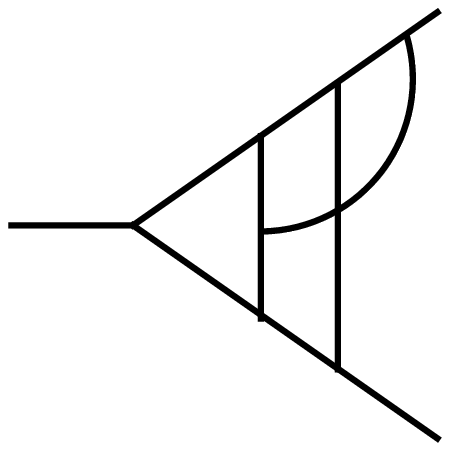} &
      \includegraphics[width=7em]{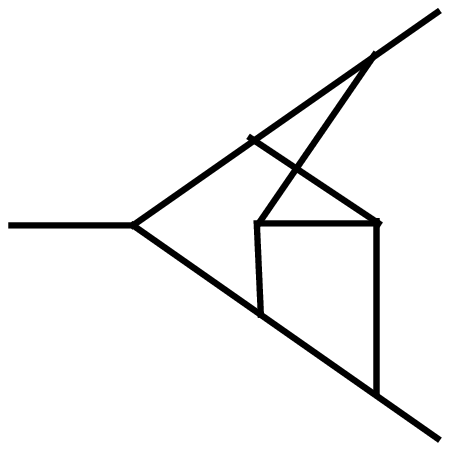}
      \\
      $A_{9,1}$ & $A_{9,2}$ & $A_{9,4}$
    \end{tabular}
    \caption{\label{fig::masters}Three most complicated master integrals
      entering the result for the three-loop form factor. The notation is
      adopted from Ref.~\cite{Gehrmann:2006wg,Heinrich:2007at}.}
  \end{center}
\end{figure}

Our results are expressed in terms of 22 master integrals. Eight master integrals
are either two-point functions or products of one- and two-loop integrals and
are thus well-known since many years (see, e.g.,
Ref.~\cite{Gonsalves:1983nq,Kazakov:1983ns,Gehrmann:2005pd,Bekavac:2005xs}).
The results for eleven three-point master integrals can be found in
Ref.~\cite{Gehrmann:2006wg,Heinrich:2007at}, however, the three most
complicated integrals, which are shown in Fig.~\ref{fig::masters},
are not yet known in the literature. Our calculation fixes, by comparing with
Ref.~\cite{Moch:2005id}, the
divergent parts of $A_{9,2}$ and $A_{9,4}$ and the finite part of
$A_{9,1}$ and leaves only three
coefficients of the $\epsilon$ expansion undetermined. The results read
(assuming massless propagators of the form $1/(k^2+i0)$ and pulling out a
factor $(i \pi^{d/2} e^{-\gamma_E \epsilon})^3 (-q^2-i0)^{-3-3\epsilon}$)
\begin{eqnarray}
  A_{9,1} &=&
  \frac{1}{18\epsilon^5}
  - \frac{1}{2\epsilon^4}
  + \frac{1}{\epsilon^3}
  \left(
  \frac{53}{18} + \frac{29\zeta(2)}{36}
  \right)
  + \frac{1}{\epsilon^2}
  \left(
  -\frac{29}{2}
  \right.\nonumber\\&&\left.\mbox{}
  - \frac{149\zeta(2)}{36} + \frac{35\zeta(3)}{18}
  \right)
  + \frac{1}{\epsilon}
  \left(
  \frac{129}{2}
  + \frac{139\zeta(2)}{12}
  \right.\nonumber\\&&\left.\mbox{}
  - \frac{307\zeta(3)}{18}
  + \frac{5473\zeta(4)}{288}
  \right)
  - \frac{537}{2}
  - \frac{57\zeta(2)}{4}
  \nonumber\\&&\mbox{}
  + \frac{1103\zeta(3)}{18}
  - \frac{15625\zeta(4)}{288}
  + \frac{871\zeta(2)\zeta(3)}{36}
  \nonumber\\&&\mbox{}
  + \frac{793\zeta(5)}{10}
  + \epsilon\,X_{9,1}
  + {\cal O}\left(\epsilon^2\right)
  \,,
  \label{eq::A91}
\end{eqnarray}
\begin{eqnarray}
  A_{9,2} &=&
  - \frac{2}{9\epsilon^6}
  - \frac{5}{6\epsilon^5}
  + \frac{1}{\epsilon^4}
  \left(
    \frac{20}{9}
  + \frac{17\zeta(2)}{9}
  \right)
  + \frac{1}{\epsilon^3}
  \left(
  - \frac{50}{9}
\right.\nonumber\\&&\left.\mbox{}
  + \frac{181\zeta(2)}{36}
  + \frac{31\zeta(3)}{3}
  \right)
  + \frac{1}{\epsilon^2}
  \left(
    \frac{110}{9}
  - \frac{34\zeta(2)}{3}
\right.\nonumber\\&&\left.\mbox{}
  + \frac{347\zeta(3)}{18}
  + \frac{595\zeta(4)}{24}
  \right)
  + \frac{1}{\epsilon}
  \left(
  - \frac{170}{9}
  + 19\zeta(2)
\right.\nonumber\\&&\left.\mbox{}
  - \frac{514\zeta(3)}{9}
  + \frac{489\zeta(4)}{32}
  - \frac{341\zeta(2)\zeta(3)}{6}
\right.\nonumber\\&&\left.\mbox{}
  + \frac{2507\zeta(5)}{15}
  \right)
  + X_{9,2}
  + {\cal O}\left(\epsilon\right)
  \,,
  \label{eq::A92}
\end{eqnarray}
\begin{eqnarray}
  A_{9,4} &=&
  - \frac{1}{9\epsilon^6}
  - \frac{8}{9\epsilon^5}
  + \frac{1}{\epsilon^4}
  \left(
  1
  + \frac{43\zeta(2)}{18}
  \right)
  + \frac{1}{\epsilon^3}
  \left(
  \frac{14}{9}
\right.\nonumber\\&&\left.\mbox{}
  + \frac{106\zeta(2)}{9}
  + \frac{109\zeta(3)}{9}
  \right)
  + \frac{1}{\epsilon^2}
  \left(
  -17
\right.\nonumber\\&&\left.\mbox{}
  - \frac{311\zeta(2)}{18}
  + \frac{608\zeta(3)}{9}
  - \frac{481\zeta(4)}{144}
  \right)
  + \frac{1}{\epsilon}
  \bigg(\left.
  84
\right.\nonumber\\&&\left.\mbox{}
  + \frac{11\zeta(2)}{3}
  - \frac{949\zeta(3)}{9}
  + \frac{425\zeta(4)}{6}
  + \frac{3463\zeta(5)}{45}
  \right.\nonumber\\&&\left.\mbox{}
  - \frac{2975\zeta(2)\zeta(3)}{18}
  \right)
  + X_{9,4}
  + {\cal O}\left(\epsilon\right)
  \,.
\end{eqnarray}
We obtained a numerical result for the coefficient $X_{9,1}$ using
the Mellin-Barnes (MB) method~\cite{MB1,MB2,MB3}, starting from the general
MB representation for the tennis court diagram of
Ref.~\cite{Bern:2005iz},
and applying the corresponding packages~\cite{MB4,MB5}.
To evaluate numerically  $X_{9,2}$ and
$X_{9,4}$  we used the program {\tt FIESTA}~\cite{FIESTA}
which is a convenient and efficient implementation of the sector
decomposition algorithm. 
Our results read
\begin{eqnarray}
  X_{9,1} \approx 1429(1)\,,
  X_{9,2} \approx 528.0(4)\,,
  X_{9,4} \approx -2085(5)\,,
  \label{eq::X9}
\end{eqnarray}
where the accuracy is sufficient for all foreseeable
physical applications.
Finally, let us mention that we evaluate the colour factors with the
help of the program {\tt color}~\cite{vanRitbergen:1998pn}.

In the following we want to present explicit results for $F_q$ and $F_g$.
We parameterize the results in terms of the bare coupling which allows us to
factorize all occurring logarithms of the form $\ln(Q^2/\mu^2)$ where
$Q^2=-q^2>0$. Furthermore, we cast the results in the form ($x=q,g$)
\begin{eqnarray}
  F_x &=& 1 + \sum_n \left(\frac{\alpha_s}{4\pi}\right)^n
  \left(\frac{\mu^2}{Q^2}\right)^{n\epsilon} F_x^{(n)}
  \,,
\end{eqnarray}
and split $F_q^{(3)}$ into the singlet, fermionic and remaining gluonic part
\begin{eqnarray}
  F_q^{(3)} &=& F_q^{(3),g} + F_q^{(3),n_f} + \sum_{q^\prime} Q_{q^\prime} F_q^{(3),sing}
  \,,
\end{eqnarray}
where $n_f$ stands for the number of active quarks.
The results for $F_q^{(1)}$ and $F_q^{(2)}$ (expanded in $\epsilon$ sufficient
for the three-loop calculation) can be found in Eqs.~(3.5) and~(3.6) of
Ref.~\cite{Moch:2005id} and $F_g^{(1)}$, $F_g^{(2)}$ and $F_q^{(3),n_f}$
are given in Eqs.~(7),~(8) and~(6) of Ref.~\cite{Moch:2005tm}, respectively.
The pole parts of $F_q^{(3),g}$ and $F_g^{(3)}$ are listed in Eqs.~(3.7) of
Ref.~\cite{Moch:2005id} and~(9) of Ref.~\cite{Moch:2005tm}, respectively.
Our expressions agree with all these results which constitutes a strong cross
check since in Refs.~\cite{Moch:2005id,Moch:2005tm} a completely different
approach has been chosen to evaluate the Feynman integrals. In particular,
no reduction to master integrals has been performed.
In this Letter new results for $F_q^{(3),g}$, $F_q^{(3),sing}$ and $F_g^{(3)}$
are presented. Since the pole parts are already available in the literature we
display only the corresponding finite parts which read in the case of a
$SU(N_c)$ colour group
\begin{widetext}
\begin{eqnarray}
  F_q^{(3),g+n_f}\Big|_{\rm fin} &=&
  C_F^3
  \left(
     \frac{26871}{8}
    -\frac{95137 \zeta(2)}{60}
    +\frac{5569 \zeta(3)}{5}
    +\frac{95375 \zeta(4)}{48}
    +\frac{30883 \zeta(2)\zeta(3)}{15}
    -\frac{16642 \zeta(5)}{5}
    +\frac{2669 (\zeta(3))^2}{3}
  \right.\nonumber\\&&\left.\mbox{}
    +\frac{1961387 \zeta(6)}{2880}
    -\frac{24 {\conepm}}{5}
    +\frac{24 {\ctwop}}{5}
    +\frac{6{\cfourp}}{5}
  \right)
  +C_A  C_F^2
  \left(
     \frac{20003431}{29160}
    +\frac{4239679 \zeta(2)}{1620}
    -\frac{121753 \zeta(3)}{30}
  \right.\nonumber\\&&\left.\mbox{}
    -\frac{11155817 \zeta(4)}{4320}
    -\frac{92554 \zeta(2)\zeta(3)}{45}
    +\frac{610462 \zeta(5)}{225}
    -\frac{36743 (\zeta(3))^2}{30}
    -\frac{1118529 \zeta(6)}{640}
    +\frac{24 {\conepm}}{5}
  \right.\nonumber\\&&\left.\mbox{}
    -\frac{16 {\ctwop}}{5}
    -\frac{9  {\cfourp}}{5}
  \right)
  +C_A^2C_F \left(
    -\frac{88822328}{32805}
    -\frac{3486997 \zeta(2)}{2916}
    +\frac{3062512  \zeta(3)}{1215}
    +\frac{4042277 \zeta(4)}{4320}
  \right.\nonumber\\&&\left.\mbox{}
    +\frac{5233 \zeta(2) \zeta(3)}{12}
    -\frac{202279 \zeta(5)}{450}
    +\frac{63043 (\zeta(3))^2}{180}
    +\frac{4741699 \zeta(6)}{11520}
    -{\conepm}
    +\frac{2 {\ctwop}}{5}
    +\frac{3 {\cfourp}}{5}
  \right)
  \nonumber\\&&\mbox{}
  +C_F^2n_fT
  \left(
    -\frac{2732173}{1458}
    -\frac{45235\zeta(2)}{81}
    +\frac{102010 \zeta(3)}{81}
    +\frac{40745 \zeta(4)}{216}
    -\frac{686 \zeta(3) \zeta(2)}{9}
    +\frac{556 \zeta(5)}{45}
  \right)
  \nonumber\\&&\mbox{}
  +C_AC_Fn_fT
  \left(
     \frac{17120104}{6561}
    +\frac{442961\zeta(2)}{729}
    -\frac{90148 \zeta(3)}{81}
    -\frac{5465\zeta(4)}{27}
    +\frac{736 \zeta(3) \zeta(2)}{9}
    -\frac{416 \zeta(5)}{3}
  \right)
  \nonumber\\&&\mbox{}
  +C_Fn_f^2
  T^2
  \left(
    -\frac{2710864}{6561}
    -\frac{248 \zeta(2)}{3}
    +\frac{12784 \zeta(3)}{243}
    -\frac{166 \zeta(4)}{27}
  \right)
  \,,
  \\
  F_q^{(3),sing}\Big|_{\rm fin} &=&
  d^{abc}d^{abc} \left(
     \frac{2}{3}
    + \frac{5 \zeta(2)}{3}
    +\frac{7 \zeta(3)}{9}
    -\frac{\zeta(4)}{6}
    -\frac{40 \zeta(5)}{9}
  \right)
  \,,\\
  F_g^{(3)}\Big|_{\rm fin} &=&
  C_A^3 \left(
     \frac{14423912}{6561}
    +\frac{384479 \zeta(2)}{2916}
    -\frac{370649 \zeta(3)}{486}
    +\frac{280069 \zeta(4)}{864}
    +\frac{1821 \zeta(2)\zeta(3)}{4}
    -\frac{66421 \zeta(5)}{90}
  \right.\nonumber\\&&\left.\mbox{}
    +\frac{545 (\zeta(3))^2}{36}
    -\frac{167695 \zeta(6)}{256}
    -{\conepm}+2{\ctwop}
  \right)
  +C_A^2n_f T
  \left(
    -\frac{10021313}{6561}
    -\frac{75736\zeta(2)}{729}
    -\frac{1508 \zeta(3)}{27}
  \right.\nonumber\\&&\left.\mbox{}
    +\frac{437 \zeta(4)}{12}
    -\frac{878 \zeta(3) \zeta(2)}{9}
    +\frac{6476 \zeta(5)}{45}
  \right)
  +C_FC_A n_f T
  \left(
    -\frac{155629}{243}
    -\frac{82 \zeta(2)}{3}
    +\frac{23584 \zeta(3)}{81}
    -16 \zeta(4)
  \right.\nonumber\\&&\left.\mbox{}
    +96 \zeta(3) \zeta(2)
    +\frac{64 \zeta(5)}{9}
  \right)
  +C_F^2n_f T
  \left(
     \frac{608}{9}
    +\frac{592 \zeta(3)}{3}
    -320 \zeta(5)
  \right)
  + C_F n_f^2 T^2
  \left(
     \frac{42248}{81}
    -\frac{64 \zeta(2)}{3}
  \right.\nonumber\\&&\left.\mbox{}
    -\frac{2816 \zeta(3)}{9}
    -\frac{224 \zeta(4)}{3}
  \right)
  +C_An_f^2 T^2
  \left(
     \frac{2958218}{6561}
    + \frac{304 \zeta(2)}{27}
    +\frac{47296 \zeta(3)}{243}
    +\frac{1594 \zeta(4)}{27}
  \right)
  \,,
\end{eqnarray}
\end{widetext}
where $C_F=(N_c^2-1)/(2N_c)$, $C_A=N_c$, $T=1/2$ and 
$d^{abc}d^{abc}=(N_c^2-1)(N_c^2-4)/N_c$.
Inserting numerical values leads to
$F_q^{(3),g+n_f}|_{\rm fin} \approx
-13656.8 + 3062.1 n_f - 164.2 n_f^2 \pm 2.2\delta_{9,1} \pm
0.4\delta_{9,2} \pm 2.2\delta_{9,4}$,
$F_q^{(3),sing}|_{\rm fin} \approx -5.944$,
and
$F_g^{(3)}|_{\rm fin} \approx
26102.7 - 8298.8 n_f + 585.3 n_f^2  \pm 27.0\delta_{9,1} \pm 21.6 \delta_{9,2}$,
where $\delta_{9,i}=1$ corresponds to the one sigma uncertainty given in Eq.~(\ref{eq::X9}).

It is interesting to specify our result to a supersymmetric Yang-Mills theory
containing a bosonic and fermionic degree of freedom in the same colour
representation. This is achieved by setting $C_A=C_F=2T$ and $n_f=1$ which leads to
\begin{eqnarray}
  \lefteqn{
    F_q^{(3),g+n_f}\Big|_{\rm fin} =
    C_A^3 \Bigg(
    \frac{389216}{243}
    -\frac{155935 \zeta(2)}{972}}
  \nonumber\\&&\mbox{}
  -\frac{54703 \zeta(3)}{162}
  +\frac{23897\zeta(4)}{72}
  +\frac{15875 \zeta(2)\zeta(3)}{36}
  \nonumber\\&&\mbox{}
  -\frac{11279 \zeta(5)}{10}
  +\frac{545 (\zeta(3))^2}{36}
  -\frac{167695\zeta(6)}{256}
  \nonumber\\&&\mbox{}
  -{\conepm}+2{\ctwop}
  \Bigg)
  \,,
\end{eqnarray}
\begin{eqnarray}
  \lefteqn{F_g^{(3)}\Big|_{\rm fin} =
  C_A^3 \Bigg(
    \frac{676219}{486}
    +\frac{61937 \zeta(2)}{972}
    -\frac{93295 \zeta(3)}{162}}
  \nonumber\\&&\mbox{}
  +\frac{95171\zeta(4)}{288}
  +\frac{16361 \zeta(2)\zeta(3)}{36}
  -\frac{1645 \zeta(5)}{2}
  \nonumber\\&&\mbox{}
  +\frac{545 (\zeta(3))^2}{36}
  -\frac{167695\zeta(6)}{256}
  -{\conepm}+2{\ctwop}
  \Bigg)
  .
\end{eqnarray}
Although we do not know three coefficients analytically, we believe that
the growth of the transcendentality level continues when going to the next
order in $\epsilon$ so that all the results are at most of
transcendentality six as was predicted in
Refs.~\cite{Kotikov:2004er,Bern:2005iz}.
It is interesting to note that these terms agree between the two form factors.

To summarize, in this Letter we compute the form factors of the photon-quark
and effective Higgs boson-gluon vertex to three-loop order within massless QCD.
Our results constitute important building blocks for a number of
physical applications. Among them are the two-jet cross section in $e^+ e^-$
collisions, the Higgs boson production in gluon fusion and the lepton pair
production in proton collisions via the Drell-Yan mechanism.
Let us stress that our result represents the first complete evaluation of
three-loop QCD corrections to a three-point function.
Our results for the coefficients of the three master integrals $A_{9,1}$,
$A_{9,2}$ and $A_{9,4}$ partially overlap with those of Ref.~\cite{HKS}
where these integrals were evaluated in a direct way.
Agreement has been found for all common coefficients.



Acknowledgements. This work is supported by DFG through SFB/TR~9 and grant
RFBR-08-02-01451.
The Feynman diagrams were drawn with the help of
{\tt Axodraw}~\cite{Vermaseren:1994je} and {\tt JaxoDraw}~\cite{Binosi:2003yf}.




\end{document}